\documentclass{elsarticle}
\usepackage{graphicx}
\usepackage{xspace}

\usepackage{amsmath,amssymb}

\usepackage[dvips]{color}

\newcommand\blue{\color{blue}}

\usepackage{latexsym,graphicx,epsfig}

\newcommand{\be}{\begin{equation}}
\newcommand{\ee}{\end{equation}}
\newcommand{\bea}{\begin{eqnarray}}
\newcommand{\eea}{\end{eqnarray}}
\newcommand{\bes}{\begin{subequations}}
\newcommand{\ees}{\end{subequations}}
\newcommand{\bear}{\begin{equation}\begin{array}}
\newcommand{\eear}[1]{\end{array}\label{#1}\end{equation}}
\newcommand{\fr}[2]{\dfrac{{ #1}}{{ #2}}}

\renewcommand{\le}{\leqslant}

\def\vep{{\varepsilon}}

\newcommand{\ggam}{\mbox{$\gamma\gamma\,$}}

{\end{list}}
\newcounter{enumct}

\begin{document}
\date{}
\title{
Higgs boson parity. Discussion of the experiment}
\author{I.~F.~Ginzburg
\\
{\it Sobolev  Institute  of  Mathematics,  Department  of  Theor.  Physics,
Prosp. Ac. Koptyug, 4, and\\ Novosibirsk State University, ul. Pirogova,
2, 630090 Novosibirsk, Russia
}}

\begin{abstract}{
Recently CMS and ATLAS announced that they had measured the  Higgs boson parity.  In this note we show that   their approach can determine this parity  only under the additional assumption  that
an extension of Standard Model  of  some special type is realized in Nature.

We show that the used approach gives no information about the Higgs boson parity when assuming most  other extensions of the Standard Model.

}

\end{abstract}

\maketitle


In  recent papers the  CMS~\cite{PexpCMS1,PexpCMS2} and
ATLAS \cite{PexpATLAS1,PexpATLAS2} collaborations announced that
they had measured the parity of Higgs boson with mass of 125 GeV.	
Based on the proposals~\cite{Ptheor1}-\cite{Ptheor3}, these collaborations study correlations in the momentum distributions
of leptons produced in  decays $h\to ZZ^*\to (\ell_1\bar{\ell}_1)(\ell_2\bar{\ell}_2)$ or $h\to WW^*\to (\ell_1\nu)(\ell_2\nu)$ with leptons $\ell=e,\,\mu$.

The SM Higgs boson is definitely P-even. The problem of measuring its parity appears only in the extended models of the Higgs sector (beyond Standard Model -- BSM).  We  subdivide
such models into two groups. The  approach of refs.~\cite{PexpCMS1}-\cite{Ptheor3}
allows us to determine parity of observed Higgs boson for the models of  the first group
and gives no information on the parity of observed Higgs boson for the  wider family of the models of  the second group. \\


{\bf In the first group of models} the Higgs boson
is single scalar, and the BSM  contains some new interactions,
perhaps with CP violation \cite{Ptheor1}-\cite{Ptheor3}.
To explain the main statement, I start with a brief description of these
models in terms  useful for me.

In the SM the interaction of Higgs boson with gauge bosons comes from
a kinetic term of the Lagrangian
 \be
 D^\mu\phi^*D_\mu\phi\,. \label{kinterm}
\ee
The electroweak symmetry breaking (EWSB) gives expansion of
$\phi$ in the form
$\phi=\fr{1}{\sqrt{2}}\begin{pmatrix}G^+\\v+h+iG^0\end{pmatrix}$, where
$v=246$~GeV is v.e.v. of the Higgs field. Besides, $G^0, G^\pm$ are
components of the Goldstone mode accumulated in the longitudinal component
of $Z,\,W$ and  omitted below. After substitution of this expansion in
\eqref{kinterm} terms $\propto v^2$
give masses of $W$ and $Z$ and terms $\propto v$ give the interactions
$hWW$ and $hZZ$. For brevity we  write down only the interaction with $Z$
(the interaction with $W$ differs by coefficients only)
\be
(\bar{g}v)\,h\,\,Z^\mu Z_\mu \,. \label{SMHg}
\ee

The BSM interactions in this model can produce   new terms in the effective Lagrangian of the form  ({\it dimension 5 operator})
\be
\Delta L=\fr{h}{v}(A\, Z^{\mu\nu}Z_{\mu\nu}+B\, \vep_{\mu\nu\alpha\beta}Z^{\alpha\beta} Z^{\mu\nu})\,.  \label{Zmunu}
\ee
(the factor $v$ in the denominator is written to have dimensionless
parameters $A$ and $B$).
The  term with $B$ describes a P-odd contribution in the observed
process. The data \cite{PexpCMS1}-\cite{PexpATLAS2} show that this contribution
is small. Together with the measured SM-similar cross sections for Higgs
observation, it allows to write that the P-odd component of Higgs is
absent or  small {\it (for this group of models)}. \\


{\bf The second group } consists of many other models, discussed now as possible candidates for
BSM physics. Their common feature is the existence of additional particles
similar to a Higgs boson -- both P-even and P-odd with their
possible mixing. For example, we consider the simplest variant of such models -- the
widely discussed two-Higgs-doublet model, 2HDM (see, e.g., \cite{TDLee}) {\it(the Higgs sector of MSSM is its particular case).}

In the 2HDM the basic Higgs doublet $\phi_1$ is supplemented by a second scalar
doublet $\phi_2$.  The kinetic term is a sum of two terms, similar to \eqref{kinterm}.
The  EWSB with standard decomposition for neutral components like the SM
case $\phi_i^0=(v_ie^{i\xi_i}+\zeta_i+i\eta_i)/\sqrt{2}$ gives four neutral
fields  $\zeta_{1,2}$, \ $\eta_{1,2}$ (where $v_{1,2}$ are v.e.v.'s of the
fields $\phi_1$ and $|v_1|^2+|v_2|^2=v^2$). One  linear combination of
$\eta_{1,2}$ gives a neutral component of the Goldstone field $G^0$,
the orthogonal linear combination of $\eta_i$ is denoted by $\tilde{\eta}$.
In the CP-conserving case a linear combination of the fields $\zeta_i$  forms
two scalar Higgses $h$ and $H$, while  $\tilde{\eta}$  describes a
P-odd Higgs $A$.

In the CP-violating case the fields $\zeta_i$ and $\tilde{\eta}$ are mixed,
forming three Higgs fields $h_a$, having no definite P-parity. We denote
the observed Higgs boson as $h_1$.
The value of P-odd/P-even mixing for $h_1$ can be either small or
large  (with limitations that appear beyond the Higgs sector). To determine
Higgs P-parity, one needs to observe (and measure) this mixing.

The interaction of $h_a$ with $Z$ comes from a kinetic term in precisely
the same way as in the SM and can be written as
\be
g^Z_{SM}\sum\limits_a \chi_a^Vh_aZ^\mu Z_\mu\,.\label{haZZ}
\ee

In this main approximation the  form of $h_aZZ$  interaction does not depend on the P-parity of $h_a$ (only the "P-even part of $h_a$" interacts with $Z$),  the terms like $B\, \vep_{\mu\nu\alpha\beta}Z^{\alpha\beta} Z^{\mu\nu})$ \eqref{Zmunu} don't appear. Therefore, {\bf the experiments \cite{PexpCMS1}-\cite{PexpATLAS2} do not
allow us to draw any conclusions about the Higgs boson parity}  if this model (and  many other models) is realized.

In the radiative corrections small terms like \eqref{Zmunu}  can appear. The CMS--ATLAS  data support this smallness. \\

{\it In order to finish our discussion, we show that -- within 2HDM -- even a big admixture of P-odd components in the observed Higgs boson does not contradict the modern data.} For this goal  we  show that,  with a suitable choice of parameters, the
same values of the  cross sections $gg\to h\to \ggam$ and
$gg\to h\to ZZ$ can be obtained both in the SM  and in the strongly CP-violating case of 2HDM with $M_1=M_h=125$~GeV,
\bear{l}
\fr{\sigma(gg\to h_1\to \ggam)}{\sigma(gg\to h\to \ggam)_{SM}} =1\,,\qquad  \fr{\sigma(gg\to h_1\to ZZ)}{\sigma(gg\to h\to ZZ)_{SM}}= 1\,.
\eear{estim}
 Some  sets of  parameter values which satisfy these equations (with taking into account some other limitations) are presented in \cite{bigCP2HDM} for the cases when the $t\bar{t}h_1$ production cross section will differ significantly from its SM value.

We are interested in the case when, in addition to \eqref{estim}, future measurements of the $t\bar{t}h_1$ production cross section will give results which are  very close to predictions of the SM.

In our calculations
 we use relative couplings, determined
for the neutral Higgs bosons $h_a$ with mass $M_a$ and for the charged
Higgs bosons $H^\pm$ with mass $M_\pm$:
\bear{c}
\chi^P_{a}=\fr{g^P_a}{g^P_{\rm SM}} \;\; (P=V\,(W,Z) , q=(t,b,...))\,,\quad\chi^\pm_a=\fr{g(H^+H^-h_a)}{2M_\pm^2/v}\,.
\eear{relcoupldef}
(The ratio $\chi_3^V/\big(\chi_1^V\sqrt{1-(\chi_2^V)^2}\;\big)$ describes the admixture of the CP odd state in $h_1$  for 2HDM \cite{GKan15}.)

Using the well-known equations for the two-photon and
two-gluon widths, collected e.g. in \cite{TDLee}, \cite{Gintriple},
we determine two benchmark sets of  parameters, giving ratios \eqref{estim} at $|\chi_1^t|=1$:
\bear{c}
(I)\;\;\begin{array}{l} \chi_1^V=0.9,\quad \chi_1^\pm=0.4,
 \quad Re(\chi_1^t)=0.9, \;\;   Im(\chi_1^t)=0.43\,;\end{array}\\[3mm]
(II)\;\; \begin{array}{l}\chi_1^V=0.8, \quad \chi_1^\pm=1.4,\quad
 Re(\chi_1^t)=0.74,\;\; Im(\chi_1^t)=0.67\,.\end{array}
\eear{bench}
(We neglected all fermion contributions except $t$-quarks.)

The couplings $\chi_a^V$ obey a sum rule $\sum (\chi_a^V)^2=1$. Therefore,
in  case (I) the sum $(\chi_2^V)^2+(\chi_3^V)^2=0.19$, which allows us to have
$\chi_2^V\approx \chi_3^V\approx 0.3$ (the admixture of the P-odd to P-even
components of the $h_1$ about 0.3).
In  case (II) the sum $(\chi_2^V)^2+(\chi_3^V)^2=0.36$, which allows us to
have $\chi_2^V\approx \chi_3^V\approx 0.4$ (the admixture of the P-odd to
P-even components of the $h_1$ about 0.5).

This simple analysis shows that a big P-odd admixture in the observed Higgs boson is compatible with SM-like values for many observed rates.

Certainly, detailed analysis of data with taking into account full equations is necessary.

{\bf Summary}.  The  results \cite{PexpCMS1}-\cite{PexpATLAS2} show definitely that all observations at the LHC are consistent with the expectations for the
standard model Higgs boson with the quantum numbers
$J^{PC} = 0^{++}$. Nevertheless, these  data give no model independent information about the parity of the observed Higgs boson.

The signal of indefinite P-parity of the Higgs boson can be obtained at the LHC relatively soon in the observations $h_1\to\tau\bar{\tau}$ \cite{Askew} and, perhaps,  in  the study of the process $pp\to t\bar{t}\,h+...$ \cite{tth}. Unfortunately, these experiments cannot provide information about the shares of the P-odd and P-even components in the observed Higgs boson (since   even in the 2HDM the Yukawa sector  is almost arbitrary for a given scalar sector). The value of $h_1WW$ coupling, measurable in the experiments with $W$ fusion, will be very important for the problem considered. If it is found that $\vep^V=(1-(\chi_1^V)^2)\ll 1$, then the  P-odd fraction of the observed Higgs will be $\le \sqrt{\vep^V}$.

To measure  the shares of the P-odd and P-even components in the
Higgs boson, one should observe two other  neutral Higgs bosons
$h_{2,\,3}$ and their couplings to gauge bosons (a very complicated experiment). Direct measurement of this mixing is possible at a photon collider (see, e.g., \cite{GinIv}) when it is constructed.
 \\

The discussions with   S. Eidelman, I. Ivanov, D. Kazakov, K.~Melnikov and M.~Vysotsky were useful.
This work was supported in part by the grants RFBR  15-02-05868,
NSh-3802.2012.2 and  NCN OPUS 2012/05/B/ST2/03306 (2012-2016).\\


\begin{thebibliography}{100}





\bibitem{PexpCMS1} The CMS Collaboration.
{\it Phys. Rev. Lett.} {\bf 110}, 081803 (2013); arXiv:1212.6639v2 [hep-ex].


\bibitem{PexpCMS2} The CMS Collaboration.
{\it Phys. Rev.} {\bf  D 92}, 012004 (2015); arXiv:1411.3441 [hep-ex].


\bibitem{PexpATLAS1} The ATLAS Collaboration.
{\it Eur. Phys. J.} {\bf C75}, 231 (2015); arXiv:1503.03643.

\bibitem{PexpATLAS2} The ATLAS Collaboration.
{\it Eur. Phys. J.} {\bf C};	arXiv:1506.05669 [hep-ex].

\bibitem{Ptheor1} S.Y. Choi, D.J. Miller, M.M. Muhlleitner, P.M. Zerwas,
{\it Phys. Lett.} {\bf B 553}, 61 (2003); arXiv:hep-ph/0210077.

\bibitem{Ptheor2} S. Bolognesi et al.,
{\it Phys. Rev.} {\bf D 86}, 095031 (2012); arXiv:1208.4018 [hep-ph].


\bibitem{Ptheor3}    C. P. Buszello, I. Fleck, P. Marquard, J. J. van der Bij,
{\it Eur. Phys. J.} {\bf C32}, 209 (2004); arXiv:hep-ph/0212396.




\bibitem{TDLee} T.D. Lee, {\it Phys. Rev.} {\bf D8}, 1226 (1973);
J.F. Gunion, H.E. Haber, G. Kane and S.~Dawson, {\em The Higgs
Hunter's Guide} (Addison-Wesley, Reading, 1990);
G.~C.~Branco, P.~M.~Ferreira, L.~Lavoura, M.~N.~Rebelo, M.~Sher
and J.~P.~Silva,
    Phys.\ Rep.\  {\bf 516}, 1 (2012).





\bibitem{bigCP2HDM} D. Fontes, J.C. Romao, J.P. Silva,
R. Santos, arXiv:1506.00860 [hep-ph].

\bibitem{GKan15} I.F. Ginzburg, K.A. Kanishev. {\it Phys. Rev.} {\bf D 92} (2015) 015024; arXiv:1502.06346 [hep-ph]


\bibitem{Gintriple} I.F. Ginzburg,
arXiv:1505.01984 [hep-ph].


\bibitem{Askew}  A. Askew, P. Jaiswa,  T. Okui, H.B. Prosper, N. Sato,
\emph{ Phys. Rev.} {\bf D 91}, 075014 (2015);  arXiv:1501.03156 [hep-ph]; S.~Berge, W.~Bernreuther, S.~Kirchner, arXiv:1510.03850 [hep-ph].


\bibitem{tth} D. Fontes, J.C. Romao, J.P. Silva, arXiv:1506.00860 [hep-ph]; H.-L. Li,  P.-C. Lu, Z.-G. Si, Y. Wang, arXiv:1508.06416 [hep-ph].

\bibitem{GinIv}  I.F. Ginzburg, I.P. Ivanov,
{\it Eur. Phys. J.\ } {\bf C 22}, 411 (2001); arXiv:hep-ph/0004069.


\end{thebibliography}
\end{document}